**Voxel-Based Conversion of Hypofractionated Radiotherapy Dose Distributions to 2 Gy-Equivalent OAR Constraints: Proof-of-Concept Demonstrating the Radiobiological Benefits of Hypofractionation in a Prostate Radiotherapy Case**

Mazen Moussallem[1, 2, 3, *], Dima Mahmoud[3], Antoine Nassif[1]

[1]Healthy Innovations, Rachiine, Zgharta, Lebanon

[2]Holy Family University, Batroun, Lebanon

[3] Tripoli Governmental Hospital and Doctors Center, Kobbe, Tripoli, Lebanon

**Corresponding author, mazenphm@hotmail.com*

*April 11, 2026*

**Voxel-Based Conversion of Hypofractionated Radiotherapy Dose Distributions to 2 Gy-Equivalent OAR Constraints: Proof-of-Concept Demonstrating the Radiobiological Benefits of Hypofractionation in a Prostate Radiotherapy Case**

**Objectives:** Existing voxel-based dose converters transform hypofractionated dose distributions into biologically effective dose (BED) or equivalent dose in 2 Gy fractions ($EQD_2$), but they are not reliably applicable to organ-at-risk (OAR) dose constraints, particularly in low-dose regions, leading to potential underestimation. This study develops and demonstrates a voxel-based method to convert hypofractionated dose distributions into 2 Gy-equivalent OAR constraints, rather than relying solely on $EQD_2$.

**Methods:** The To2GyConstraints converter (Radiotherapy TPS Extensions, Healthy Innovations, www.healthy-innovations.com) was applied for radiobiological evaluation of a prostate cancer case to transform hypofractionated doses into 2 Gy-equivalent OAR constraints. This tool is based on the Linear Quadratic (LQ) model for doses per fraction ≤ 7.5 Gy and the Linear Quadratic Linear (LQ-L) model for higher doses. For voxel fraction doses below a threshold defined as the mean between the prescribed hypofractionated fraction dose and 2 Gy, an equivalent number of fractions is calculated. The method then applies an EQDx-type conversion, rather than $EQD_2$, using this calculated fraction number to better reproduce normofractionated dose behavior.

**Results:** For doses above the defined threshold, unlike BED, the To2GyConstraints model produced results consistent with $EQD_2$ and provided clinically realistic dose values comparable to standard dosimetric constraints, thereby offering a clearer demonstration of the radiobiological benefits of hypofractionation in prostate cancer. For doses below the threshold, unlike $EQD_2$, the To2GyConstraints model showed behavior consistent with BED, yielding higher dose estimates when converted to a normofractionation scheme.

**Conclusions:** This converter shows promising results but requires multicenter validation across multiple disease sites and large cohorts, including comparing hypofractionated constraint violations with those of normofractionation after dosimetric conversion using this tool.

**Advances in knowledge:** This study introduces a voxel-based approach that enables the application of normofractionation OAR dose constraints to hypofractionated dosimetry after conversion.

## INTRODUCTION

Reirradiation is increasingly relevant in modern radiotherapy and requires high precision to deliver an adequate tumor dose while avoiding exceeding tolerance doses in organs at risk (OARs) that have already received radiation [1, 2]. This process also requires dose evaluation in 2 Gy fractions, as a consistent fractionation scheme is needed to enable reliable dose summation across treatment plans. When a previous dosimetric plan is combined with a new plan using a different fractionation schedule, there is no clear or standardized guidance on which fractionation-based OAR constraints should be applied to evaluate the cumulative dose. For this reason, converting both plans into a form compatible with a normofractionation (2Gy per fraction) framework is of great interest, as normofractionation is the classical regimen for which extensive clinical trial data and well-established OAR constraints are available.

There are some existing software tools that perform dose distribution conversion to a normofractionation framework [3-6]; however, these tools lack clinical validation for OARs dose constraints. In addition, they are not suitable for doses per fraction greater than 7.5 Gy, which corresponds to the limit of the Linear Quadratic (LQ) model. Furthermore, these tools are generally based on converting voxel doses either into biologically effective dose (BED), which does not provide clinically realistic values comparable to standard dosimetric constraints, or into equivalent dose in 2 Gy fractions ($EQD_2$) by converting all voxel doses to a 2 Gy-per-fraction scheme, which does not accurately represent true normofractionation dose distributions.

In clinical practice, dose constraints are derived from spatial dose distributions characterized by approximately 2 Gy per fraction in the tumor region, with decreasing dose per fraction as the distance from the target increases. This spatial behavior is not reflected in current software approaches, which uniformly convert all voxels to a 2 Gy-per-fraction scheme while implicitly adjusting the number of fractions to maintain a fixed dose per fraction of 2 Gy. This approach may lead to underestimation of the cumulative normofractionated equivalent dose, particularly in low-dose regions. For example, voxels located far from the tumor that receive less than 2 Gy per fraction in hypofractionated plans may have their total dose reduced when converted to a 2 Gy-per-fraction framework, which is not physically or clinically appropriate, as it should not result in a reduction of total dose when converting to a normofractionation scheme. As a result, these methods implicitly assume that all organs are located within or very close to the planning target volume (PTV), which is not representative of true clinical normofractionation dosimetry, for which dose constraints were originally established.

To overcome the current limitations, the present study developed a voxel-based conversion method for transforming hypofractionated dose distributions into 2 Gy-equivalent OAR constraints. The method was tested on a single prostate cancer case to provide an initial assessment of its performance and to clearly illustrate the radiobiological benefits of hypofractionation in prostate radiotherapy.

## METHODS AND MATERIALS

### Description of the To2GyConstraints Model

The To2GyConstraints version 1.0 converter is one of the software modules available within the Radiotherapy TPS Extensions product line developed by Healthy Innovations (www.healthy-innovations.com). It allows users to upload radiotherapy Digital Imaging and Communications in Medicine (DICOM RT) files from a treatment plan, specify the α/β value for each organ, which is a radiobiological parameter that characterizes tissue sensitivity to dose per fraction in radiotherapy [7–11], and automatically receive the converted files for import into their local treatment planning system (TPS). For all voxels without a user-defined α/β value, a default value of 4 Gy is automatically assigned.

This converter processes all voxels within the dose distribution by converting the dose in each voxel to its equivalent value in a normofractionated dosimetry framework. The resulting converted distribution can then be used to evaluate hypofractionated dosimetry through its normofractionated equivalent, using conventional dose–volume histograms (DVHs) and standard OAR constraints established for normofractionated treatment plans.

In To2GyConstraints version 1.0, used in the present study, a transition dose ($D_t$) of 7.5 Gy was predefined by the developers. This threshold guides the software to apply conversion based on the LQ model to the LQ model (LQ → LQ) when the voxel dose per fraction is less than $D_t$. Conversely, when the voxel dose per fraction exceeds $D_t$, conversion is performed from the Linear Quadratic-Linear (LQ-L) model to the LQ model (LQ-L → LQ). This choice was made because the LQ model is more appropriate for doses per fraction below $D_t$, whereas the LQ-L model is more suitable for doses per fraction above $D_t$ [7]. Consequently, for each voxel, dose conversion from state 1 to state 2 is performed according to the following definitions:

- $n_i$: number of fraction in state i
- $d_i$: dose per fraction (Gy) in state i
- $D_i$: cumulative dose (Gy) in state i ($D_i = n_i \times d_i$)
- $BED_{LQ}(n_i, d_i)$: biologically effective dose according to the LQ model in state i, defined as: $\mathrm{BED_{LQ}}(\mathrm{n_i}, \mathrm{d_i}) = n_i d_i \left(1 + \frac{d_i}{\alpha/\beta}\right)$, for $d_i \leq D_t$
- $BED_{LQ\text{-}L}(n_i, d_i)$: biologically effective dose according to the LQ-L model in state i, defined as: $\mathrm{BED_{LQ-L}}(\mathrm{n_i}, \mathrm{d_i}) = n_i d_i \left(1 + \frac{2\,D_t}{\alpha/\beta}\right) - \frac{n_i D_t^2}{\alpha/\beta}$, for $d_i > D_t$
- For LQ → LQ, the transition from state 1 to state 2 is obtained by solving the following equation: $\mathrm{BED_{LQ}}(\mathrm{n_1}, \mathrm{d_1}) = \mathrm{BED_{LQ}}(\mathrm{n_2}, \mathrm{d_2})$

- For LQ-L → LQ, the transition from state 1 to state 2 is obtained by solving the following equation: $\mathrm{BED_{LQ-L}}(\mathrm{n_1}, \mathrm{d_1}) = \mathrm{BED_{LQ}}(\mathrm{n_2}, \mathrm{d_2})$
- When $d_2$ = 2 Gy in $BED_{LQ}(n_2, d_2)$, $D_2$ (where $D_2 = n_2 \times d_2$) corresponds toEQD$_2$. When $d_2$ ≠2 Gy, $D_2$ can be expressed as $EQD_x$.
- An automatic $EQD_x$ calculator is available as a software module on www.healthy-innovations.com.

Before performing any dose conversion, the converter defines a dose threshold as the mean between the prescribed hypofractionated fraction dose and 2 Gy. For each voxel, if the dose per fraction is less than or equal to the minimum of this threshold and $D_t$, the voxel dose is converted to normofractionation by calculating its $EQD_x$. Otherwise, the voxel dose is converted to normofractionation using $EQD_2$. When $EQD_x$ is required, it cannot be computed without determining the number of fractions in state 2 ($n_2$). For each voxel, $n_2$ is obtained by calculating the $EQD_2$ using the defined threshold as $d_1$ and the number of fractions of the hypofractionated prescription as $n_1$, together with the α/β value of the organ in which the voxel is located. In regions where an organ overlaps with another structure, the highest dose is assigned to each voxel within the intersection, corresponding to the lowest α/β value.

## Application to a Prostate Cancer Case

To test the converter and clearly illustrate the benefits of hypofractionation in prostate cancer, a 60 Gy/20 fractions dosimetric plan was converted to normofractionation using To2GyConstraints version 1.0. α/β values were chosen as in Singh et al. [5]: 1.4, 3, and 3 Gy for the prostate, rectum, and bladder, respectively. Since the PTV includes the prostate as well as portions of the rectum and bladder, overlapping regions are present between these structures. As mentioned above, in regions where an organ overlaps another structure, the highest dose is assigned. Therefore, a new structure named PTV-OAR was created prior to conversion. This was done to avoid overriding the α/β values of the rectum and bladder regions included within the original PTV with the lower α/β value of the prostate. After converting the DICOM RT files, they were re-imported into the Monaco TPS (version 5.11.03), and a comparison was performed between the hypofractionated and normofractionated dose distributions. In addition, an axial CT slice on which the three organs could be visualized was selected, and four points of interest were placed on this slice as follows:

- The first point in the prostate, receiving 60 Gy in the hypofractionated dosimetry
- The second point in the rectal portion within the PTV, receiving 60 Gy in the hypofractionated dosimetry
- The third point in the bladder portion within the PTV, receiving 60 Gy in the hypofractionated dosimetry
- The fourth point on the 30 Gy isodose line in the hypofractionated dosimetry

Subsequently, the dose at each point was converted using BED, EQD2, and To2GyConstraints, followed by a comparison of the resulting values.

## RESULTS

Table 1 clearly shows that the first three points of interest located within the PTV were selected to receive 60 Gy in the hypofractionated scheme. However, when these doses are converted to normofractionation using any of the three methods (BED, $EQD_2$, or To2GyConstraints), radiobiological differences appear between the dose received by the prostate and those received by the OARs (rectum and bladder). These differences are also visually illustrated in Figure 1, where the hypofractionated prostate case is converted using the To2GyConstraints converter.

In addition, Table 1 shows that dose values in the PTV obtained using the BED method (120 and 189 Gy) are far from the doses used in clinical prostate treatment planning, which are around 78 Gy. On the other hand, for the point of interest located on the 50% isodose line of the hypofractionated prescription (30 Gy), when converted to normofractionation using $EQD_2$, it yields a lower dose value of 28 Gy.

**Table 1.** Dose Variation According to Initial Hypofractionated Dose, Point of Interest Location ($\alpha/\beta$), and Dose Conversion Model Used

| Point of Interest Location | $\alpha/\beta$ (Gy) | Hypofractionated Dose in 20 Fractions (Gy) | Converted Using BED (Gy) | Converted Using $EQD_2$ (Gy) | Converted Using To2GyConstraints (Gy) |
|---|---|---|---|---|---|
| Prostate PTV-OAR | 1.4 | 60 | 189 | 78 | 78 |
| Rectal Portion Within the PTV | 3 | 60 | 120 | 72 | 72 |
| Bladder Portion Within the PTV | 3 | 60 | 120 | 72 | 72 |
| 50% Isodose Line of the Hypofractionated Prescription | 4 | 30 | 41 | 28 | 32 |

**Figure 1.** On the left, hypofractionated prostate dose distribution images corresponding to 60 Gy in 20 fractions; on the right, the converted dose distribution images obtained using the To2GyConstraints converter. The 100% colorwash dose level was set to 60 Gy on the left and 78 Gy on the right, corresponding to the EQD2 of 60 Gy for the prostate (α/β=1.4 Gy). In all images, the maximum display threshold was set to 107%. The minimum thresholds were set to 95%, 50%, and 0% for the upper, middle, and lower images, respectively. The black, red, green, and yellow crosses represent points of interest located in the prostate, the rectal portion within the PTV, the bladder portion within the PTV, and on the 50% isodose line of the hypofractionated prescription dose, respectively.

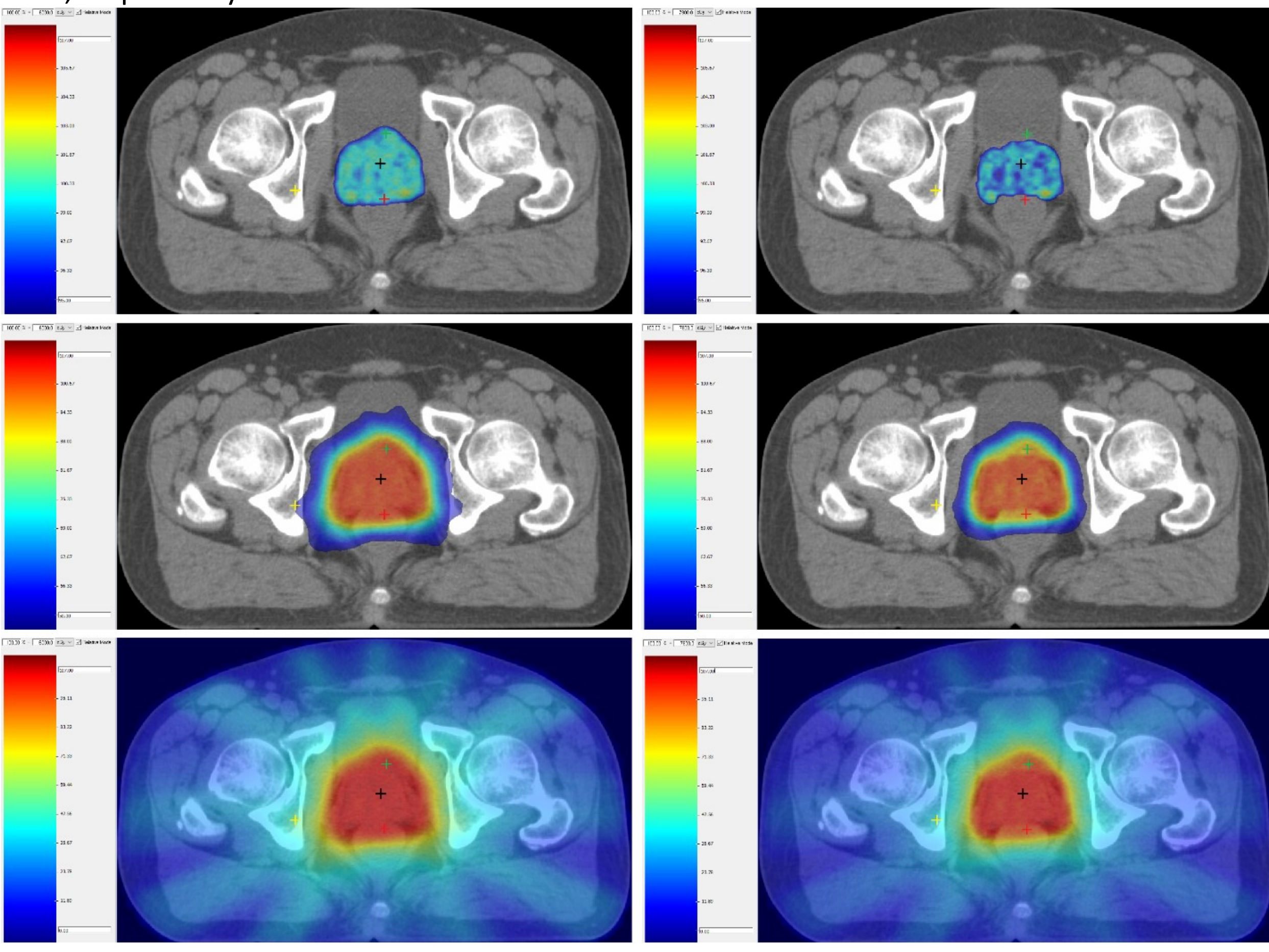

## DISCUSSION

Our results clearly demonstrate the benefit of hypofractionation in prostate cancer. We have shown that, for the prostate, which is characterized by a low α/β value, hypofractionation allows better protection of the portions of the rectum and bladder within the PTV, despite these regions receiving the same prescribed dose in the hypofractionated scheme. However, when directly applying a normofractionation scheme, this advantage is lost, and all structures within the PTV receive the same prescribed dose. A visual representation of this benefit is shown in Figure 1, which was not clearly highlighted in similar studies such as Singh et al. [3,5].

In addition, the present paper demonstrates major advantages of the To2GyConstraints approach compared with existing normofractionation converters, which are affected by the following limitations. First, the BED method yields dose values far from those used clinically for OAR constraints. Second, due to the issues detailed in the Introduction section, the $EQD_2$ method provides, for regions far from the PTV, a normofractionated dose lower than the hypofractionated dose (e.g., 28 Gy compared with 30 Gy in Table 1), as can also be seen in the colorwash images in Singh et al. [3,5], which is not radiobiologically consistent.

## CONCLUSIONS

The To2GyConstraints converter shows promising results and overcomes several limitations of existing approaches for the clinical conversion of hypofractionated schemes to normofractionated equivalents. A multicentric validation on large cohorts is recommended. In addition, this study supports the radiobiological benefit of hypofractionated radiotherapy in prostate cancer, particularly in terms of improved protection of OARs.